\pdfoutput=1

\documentclass[12pt,a4paper]{article}

\usepackage{ifthen} 
\newboolean{pdflatex}
\setboolean{pdflatex}{true} 

\newboolean{articletitles}
\setboolean{articletitles}{true} 

\newboolean{uprightparticles}
\setboolean{uprightparticles}{false} 

\def\paperauthors{LHCb collaboration} 
\def\paperasciititle{Precision measurement of the Lambdac, Xic and Xic0 baryon lifetimes} 
\def\papertitle{Precision measurement of the $\Lc$, $\Xicp$ and $\Xicz$ baryon lifetimes} 
\def\paperkeywords{{High Energy Physics}, {LHCb}} 
\def\papercopyright{\the\year\ CERN for the benefit of the LHCb collaboration} 
\def\paperlicence{CC-BY-4.0 licence}
\def\paperlicenceurl{https://creativecommons.org/licenses/by/4.0/}

\usepackage[top=1in, bottom=1.25in, left=1in, right=1in]{geometry}

\usepackage{microtype}
\usepackage{lineno}  
\usepackage{xspace} 
\usepackage{caption} 

\usepackage{graphicx}  
\usepackage{color}
\usepackage{colortbl}
\graphicspath{{./figs/}} 
\DeclareGraphicsExtensions{.pdf,.PDF,png,.PNG}

\usepackage{amsmath} 
\usepackage{amssymb}
\usepackage{amsfonts}
\usepackage{upgreek} 

\newcommand*\patchAmsMathEnvironmentForLineno[1]{%
\expandafter\let\csname old#1\expandafter\endcsname\csname #1\endcsname
\expandafter\let\csname oldend#1\expandafter\endcsname\csname
end#1\endcsname
 \renewenvironment{#1}%
   {\linenomath\csname old#1\endcsname}%
   {\csname oldend#1\endcsname\endlinenomath}%
}
\newcommand*\patchBothAmsMathEnvironmentsForLineno[1]{%
  \patchAmsMathEnvironmentForLineno{#1}%
  \patchAmsMathEnvironmentForLineno{#1*}%
}
\AtBeginDocument{%
\patchBothAmsMathEnvironmentsForLineno{equation}%
\patchBothAmsMathEnvironmentsForLineno{align}%
\patchBothAmsMathEnvironmentsForLineno{flalign}%
\patchBothAmsMathEnvironmentsForLineno{alignat}%
\patchBothAmsMathEnvironmentsForLineno{gather}%
\patchBothAmsMathEnvironmentsForLineno{multline}%
\patchBothAmsMathEnvironmentsForLineno{eqnarray}%
}

\usepackage{hyperxmp}

\usepackage[pdftex,
            pdfauthor={\paperauthors},
            pdftitle={\paperasciititle},
            pdfkeywords={\paperkeywords},
            pdfcopyright={Copyright (C) \papercopyright},
            pdflicenseurl={\paperlicenceurl}]{hyperref}

\usepackage[all]{hypcap} 

\usepackage{xspace} 
\usepackage{upgreek}

\def\lhcb   {\mbox{LHCb}\xspace}

\def\MagUp {\mbox{\em Mag\kern -0.05em Up}\xspace}

\ifthenelse{\boolean{uprightparticles}}%
{

 \def\Pmu         {\ensuremath{\upmu}\xspace}                 
 \def\Pnu         {\ensuremath{\upnu}\xspace}                 
                  
 \def\Ppi         {\ensuremath{\uppi}\xspace}

 \def\Ptau        {\ensuremath{\uptau}\xspace}

 \def\PDelta      {\ensuremath{\Delta}\xspace}                 
 \def\PXi         {\ensuremath{\Xi}\xspace}                 
 \def\PLambda     {\ensuremath{\Lambda}\xspace}                 
 \def\PSigma      {\ensuremath{\Sigma}\xspace}                 
 \def\POmega      {\ensuremath{\Omega}\xspace}                 
 \def\PUpsilon    {\ensuremath{\Upsilon}\xspace}

 \def\PB      {\ensuremath{\mathrm{B}}\xspace}                 
                  
 \def\PD      {\ensuremath{\mathrm{D}}\xspace}

 \def\PK      {\ensuremath{\mathrm{K}}\xspace}

 \def\Pb      {\ensuremath{\mathrm{b}}\xspace}                 
 \def\Pc      {\ensuremath{\mathrm{c}}\xspace}

 \def\Pi      {\ensuremath{\mathrm{i}}\xspace}

 \def\Ps      {\ensuremath{\mathrm{s}}\xspace}

 \def\thebaroffset{0.0em}
}
{

 \def\Pmu         {\ensuremath{\mu}\xspace}                 
 \def\Pnu         {\ensuremath{\nu}\xspace}                 
                  
 \def\Ppi         {\ensuremath{\pi}\xspace}

 \def\Ptau        {\ensuremath{\tau}\xspace}

 \mathchardef\PDelta="7101
 \mathchardef\PXi="7104
 \mathchardef\PLambda="7103
 \mathchardef\PSigma="7106
 \mathchardef\POmega="710A
 \mathchardef\PUpsilon="7107
                  
 \def\PB      {\ensuremath{B}\xspace}                 
                  
 \def\PD      {\ensuremath{D}\xspace}

 \def\PK      {\ensuremath{K}\xspace}

 \def\Pb      {\ensuremath{b}\xspace}                 
 \def\Pc      {\ensuremath{c}\xspace}

 \def\Pi      {\ensuremath{i}\xspace}

 \def\Ps      {\ensuremath{s}\xspace}

 \def\thebaroffset{0.18em}
}
\newcommand{\offsetoverline}[2][\thebaroffset]{\kern #1\overline{\kern -#1 #2}}%

\makeatletter
\ifcase \@ptsize \relax
  \newcommand{\miniscule}{\@setfontsize\miniscule{4}{5}}
\or
  \newcommand{\miniscule}{\@setfontsize\miniscule{5}{6}}
\or
  \newcommand{\miniscule}{\@setfontsize\miniscule{5}{6}}
\fi
\makeatother

\DeclareRobustCommand{\optbar}[1]{\shortstack{{\miniscule (\rule[.5ex]{1.25em}{.18mm})}
  \\ [-.7ex] $#1$}}

\def\mun        {{\ensuremath{\Pmu^-}}\xspace} 

\def\taum       {{\ensuremath{\Ptau^-}}\xspace}

\def\neu        {{\ensuremath{\Pnu}}\xspace}
\def\neub       {{\ensuremath{\overline{\Pnu}}}\xspace}

\def\neumb      {{\ensuremath{\neub_\mu}}\xspace}
\def\neut       {{\ensuremath{\neu_\tau}}\xspace}
\def\neutb      {{\ensuremath{\neub_\tau}}\xspace}

\def\squark    {{\ensuremath{\Ps}}\xspace}

\def\cquark    {{\ensuremath{\Pc}}\xspace}

\def\bquark    {{\ensuremath{\Pb}}\xspace}

\def\pion   {{\ensuremath{\Ppi}}\xspace}

\def\pip    {{\ensuremath{\pion^+}}\xspace}
\def\pim    {{\ensuremath{\pion^-}}\xspace}

\def\kaon    {{\ensuremath{\PK}}\xspace}

\def\KorKbar {\kern \thebaroffset\optbar{\kern -\thebaroffset \PK}{}\xspace}

\def\Kp      {{\ensuremath{\kaon^+}}\xspace}
\def\Km      {{\ensuremath{\kaon^-}}\xspace}

\def\Dbar    {{\ensuremath{\offsetoverline{\PD}}}\xspace}
\def\D       {{\ensuremath{\PD}}\xspace}

\def\DorDbar {\kern \thebaroffset\optbar{\kern -\thebaroffset \PD}\xspace}
\def\Dz      {{\ensuremath{\D^0}}\xspace}
\def\Dzb     {{\ensuremath{\Dbar{}^0}}\xspace}
\def\Dp      {{\ensuremath{\D^+}}\xspace}
\def\Dm      {{\ensuremath{\D^-}}\xspace}

\def\Dstarp  {{\ensuremath{\D^{*+}}}\xspace}

\def\Ds      {{\ensuremath{\D^+_\squark}}\xspace}
\def\Dsp     {{\ensuremath{\D^+_\squark}}\xspace}
\def\Dsm     {{\ensuremath{\D^-_\squark}}\xspace}

\def\B       {{\ensuremath{\PB}}\xspace}

\def\BorBbar {\kern \thebaroffset\optbar{\kern -\thebaroffset \PB}\xspace}

\def\Bub     {{\ensuremath{\B^-}}\xspace}

\def\Bm      {{\ensuremath{\Bub}}\xspace}

\def\Y#1S{\ensuremath{\PUpsilon{(#1S)}}\xspace}

\def\Lz          {{\ensuremath{\PLambda}}\xspace}

\def\LorLbar     {\kern \thebaroffset\optbar{\kern -\thebaroffset \PLambda}\xspace}

\def\Xires       {{\ensuremath{\PXi}}\xspace}

\def\Omegares    {{\ensuremath{\POmega}}\xspace}

\def\Lc          {{\ensuremath{\Lz^+_\cquark}}\xspace}

\def\Xicz        {{\ensuremath{\Xires^0_\cquark}}\xspace}
\def\Xicp        {{\ensuremath{\Xires^+_\cquark}}\xspace}

\def\Omegac      {{\ensuremath{\Omegares^0_\cquark}}\xspace}

\def\Lb           {{\ensuremath{\Lz^0_\bquark}}\xspace}

\def\Xibz         {{\ensuremath{\Xires^0_\bquark}}\xspace}
\def\Xibm         {{\ensuremath{\Xires^-_\bquark}}\xspace}

\def\to                 {\ensuremath{\rightarrow}\xspace}

\def\AT#1     {\ensuremath{A_{\mathrm{T}}^{#1}}\xspace}           

\def\C#1      {\ensuremath{\mathcal{C}_{#1}}\xspace}                       
\def\Cp#1     {\ensuremath{\mathcal{C}_{#1}^{'}}\xspace}                    
\def\Ceff#1   {\ensuremath{\mathcal{C}_{#1}^{\mathrm{(eff)}}}\xspace}        
\def\Cpeff#1  {\ensuremath{\mathcal{C}_{#1}^{'\mathrm{(eff)}}}\xspace}       
\def\Ope#1    {\ensuremath{\mathcal{O}_{#1}}\xspace}                       
\def\Opep#1   {\ensuremath{\mathcal{O}_{#1}^{'}}\xspace}                    

\newcommand{\nospaceunit}[1]{\ensuremath{\text{#1}}}       
\newcommand{\aunit}[1]{\ensuremath{\text{\,#1}}}       

\newcommand{\tev}{\aunit{Te\kern -0.1em V}\xspace}
\newcommand{\gev}{\aunit{Ge\kern -0.1em V}\xspace}
\newcommand{\mev}{\aunit{Me\kern -0.1em V}\xspace}
\newcommand{\kev}{\aunit{ke\kern -0.1em V}\xspace}
\newcommand{\ev}{\aunit{e\kern -0.1em V}\xspace}
\newcommand{\mevc}{\ensuremath{\aunit{Me\kern -0.1em V\!/}c}\xspace}
\newcommand{\gevc}{\ensuremath{\aunit{Ge\kern -0.1em V\!/}c}\xspace}
\newcommand{\mevcc}{\ensuremath{\aunit{Me\kern -0.1em V\!/}c^2}\xspace}
\newcommand{\gevcc}{\ensuremath{\aunit{Ge\kern -0.1em V\!/}c^2}\xspace}

\def\mum  {\ensuremath{\,\upmu\nospaceunit{m}}\xspace}

\def\fb   {\ensuremath{\aunit{fb}}\xspace}
\def\invfb   {\ensuremath{\fb^{-1}}\xspace}

\newcommand{\stat}{\aunit{(stat)}\xspace}

\newcommand{\chisq}{\ensuremath{\chi^2}\xspace}

\newcommand{\chisqip}{\ensuremath{\chi^2_{\text{IP}}}\xspace}

\def\gsim{{~\raise.15em\hbox{$>$}\kern-.85em
          \lower.35em\hbox{$\sim$}~}\xspace}
\def\lsim{{~\raise.15em\hbox{$<$}\kern-.85em
          \lower.35em\hbox{$\sim$}~}\xspace}

\def\sPlot{\mbox{\em sPlot}\xspace}

\def\pt         {\ensuremath{p_{\mathrm{T}}}\xspace}

\def\ptot       {\ensuremath{p}\xspace}

\def\evtgen     {\mbox{\textsc{EvtGen}}\xspace}

\def\geant      {\mbox{\textsc{Geant4}}\xspace}

\def\photos     {\mbox{\textsc{Photos}}\xspace}

\def\pythia     {\mbox{\textsc{Pythia}}\xspace}

\xspace

\def\tell1  {TELL1\xspace}
\def\ukl1   {UKL1\xspace}

\usepackage{cite} 
\usepackage{mciteplus}

\usepackage{longtable} 

\begin{document}

\renewcommand{\thefootnote}{\fnsymbol{footnote}}
\setcounter{footnote}{1}


\begin{titlepage}
\pagenumbering{roman}

\vspace*{-1.5cm}
\centerline{\large EUROPEAN ORGANIZATION FOR NUCLEAR RESEARCH (CERN)}
\vspace*{1.5cm}
\noindent
\begin{tabular*}{\linewidth}{lc@{\extracolsep{\fill}}r@{\extracolsep{0pt}}}
\ifthenelse{\boolean{pdflatex}}
{\vspace*{-1.5cm}\mbox{\!\!\!\includegraphics[width=.14\textwidth]{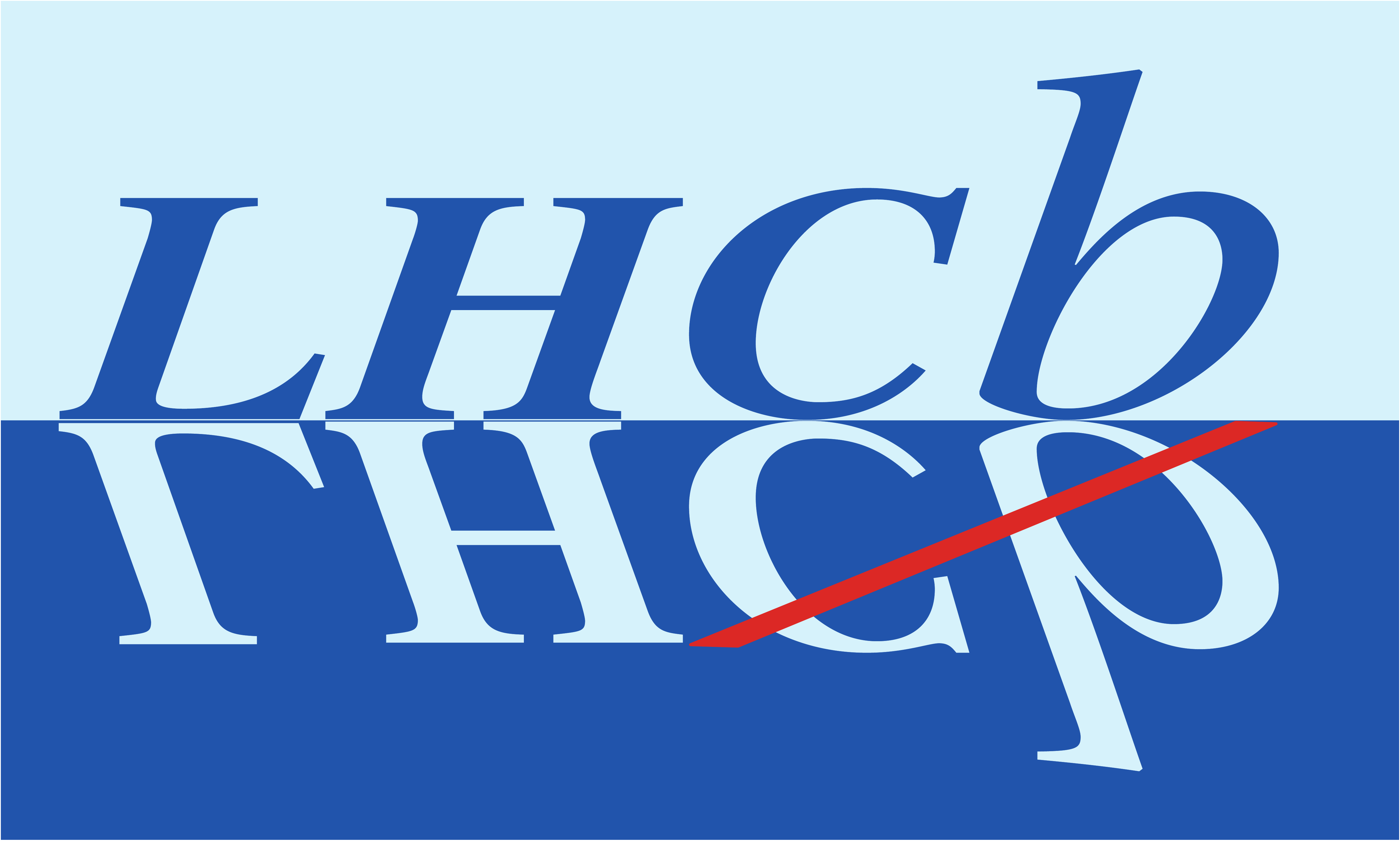}} & &}%
{\vspace*{-1.2cm}\mbox{\!\!\!\includegraphics[width=.12\textwidth]{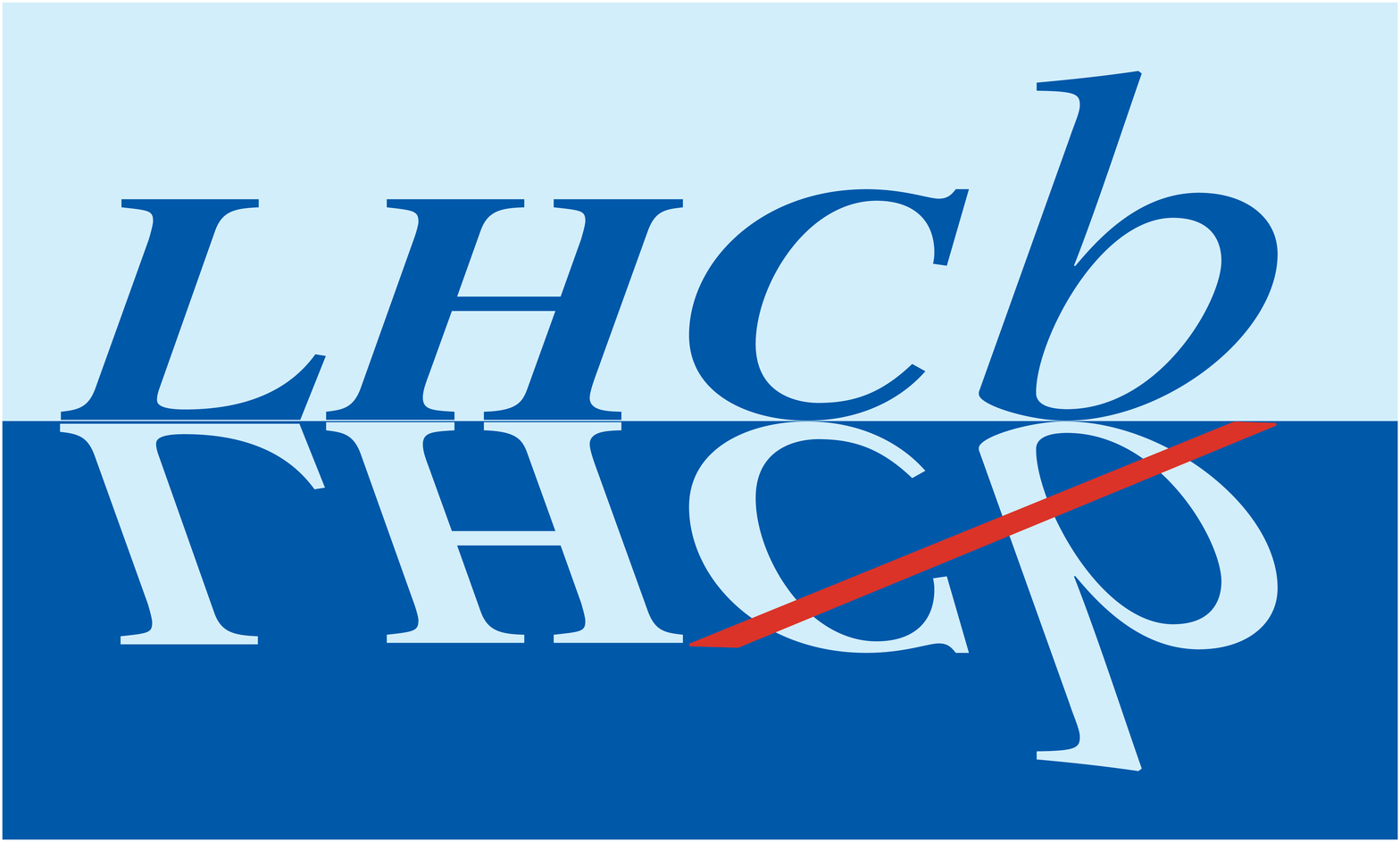}} & &}%
\\
 & & CERN-EP-2019-122 \\  
 & & LHCb-PAPER-2019-008 \\  
 & & 02 August 2019 \\ 
 & & \\
\end{tabular*}

\vspace*{2.0cm}

{\normalfont\bfseries\boldmath\huge
\begin{center}
  \papertitle 
\end{center}
}

\vspace*{2.0cm}

\begin{center}
\paperauthors\footnote{Authors are listed at the end of this paper.}
\end{center}

\vspace{\fill}

\begin{abstract}
  \noindent
 We report measurements of the lifetimes of the $\Lc$, $\Xicp$ and $\Xicz$ charm baryons
using proton-proton collision data at center-of-mass energies of 7 and 8\tev, corresponding to an 
integrated luminosity of 3.0\invfb, collected by the LHCb experiment. 
The charm baryons are reconstructed through the decays $\Lc\to p\Km\pip$, $\Xicp\to p\Km\pip$ and
$\Xicz\to p\Km\Km\pip$, and originate from semimuonic decays of beauty baryons.
The lifetimes are measured relative to that of the $\Dp$ meson, and are determined to be
\begin{align*}
  \tau_{\Lc} &= 203.5\pm1.0\pm1.3\pm1.4~{\rm fs}, \\
  \tau_{\Xicp} &= 456.8\pm3.5\pm2.9\pm3.1~{\rm fs}, \\
  \tau_{\Xicz} &= 154.5\pm1.7\pm1.6\pm1.0~{\rm fs},
\end{align*}
\noindent where the uncertainties are statistical, systematic, and due to the uncertainty in the
$\Dp$ lifetime. The measurements are approximately 3--4 times more precise than the current world 
average values. The $\Lc$ and $\Xicp$ lifetimes are in agreement with previous measurements;
however, the $\Xicz$ baryon lifetime is approximately 3.3 standard deviations larger than the world average value.

\end{abstract}

\vspace*{2.0cm}

\begin{center}
  Published in Phys. Rev. D100 (2019) 032001
\end{center}

\vspace{\fill}

{\footnotesize 
\centerline{\copyright~\papercopyright. \href{\paperlicenceurl}{\paperlicence}.}}
\vspace*{2mm}

\end{titlepage}


\newpage
\setcounter{page}{2}
\mbox{~}

\cleardoublepage


\renewcommand{\thefootnote}{\arabic{footnote}}
\setcounter{footnote}{0}



\pagestyle{plain} 
\setcounter{page}{1}
\pagenumbering{arabic}


%

Measurements of the lifetimes of hadrons containing heavy ($b$ or $c$) quarks play an important role in testing theoretical approaches
that are used to perform Standard Model calculations. The validation of such tools is important, as they
can then be used to search for deviations from Standard Model expectations in other processes. One of the most predictive tools in quark flavor physics
is the heavy quark expansion (HQE)~\cite{Khoze:1983yp,Bigi:1991ir,Bigi:1992su,Blok:1992hw,*Blok:1992he,Neubert:1997gu,Uraltsev:1998bk,Bigi:1995jr}, which
can be used to calculate the decay widths of hadrons containing heavy quarks, $Q$, through an expansion in inverse powers of 
the heavy quark mass, $m_Q$. The lowest-order term in the expansion depends only on $m_Q$, and therefore contributes equally to the
decay width of all hadrons with a single heavy quark $Q$.
Higher-order terms in the HQE are related to non-perturbative corrections, and to effects due to the 
presence of the other light (spectator) quark(s) in the heavy hadron. These corrections generally increase as the mass of the heavy quark decreases, and
therefore measurements of charm-hadron lifetimes are sensitive to these higher-order 
contributions~\cite{Kirk:2017juj,Cheng:2015iom,Lenz:2013aua,Bianco:2003vb,Bellini:1996ra,Blok:1991st}.

Particle lifetimes are also required to compare measured $b$- or $c$-hadron decay branching fractions to 
corresponding predictions for partial decay widths. Improved precision on the lifetimes thus allows for 
more stringent tests of theoretical predictions. Lastly, improving the knowledge of the properties of all Standard Model particles
is important, as they serve as input directly, or through simulation, into a wide variety of studies both within and 
beyond the Standard Model.

Recently, the LHCb collaboration reported a measurement of the $\Omegac$ lifetime~\cite{LHCb-PAPER-2018-028} that was nearly four times 
larger than, and inconsistent with, the world average value. The lifetimes of the other three ground state singly charmed baryons 
($\Lc$, $\Xicp$ and $\Xicz$) were last measured 
almost twenty years ago, and are only known with precisions of 3\%, 6\% and 10\%, respectively~\cite{PDG2018}. 
The most precise measurements contributing to the average lifetimes are those from the FOCUS 
collaboration~\cite{Link:2002ge,Link:2001qy,Link:2002xu} based on signal sample sizes of approximately 8000 $\Lc$, 500 $\Xicp$ and 100 $\Xicz$ decays. 
For the $\Lc$ baryon, there is mild tension between the average lifetime obtained from fixed target experiments~\cite{Link:2002ge,Kushnirenko:2000ed,Frabetti:1992jx} 
and that obtained by the CLEO collaboration~\cite{Mahmood:2000tw}. 

The LHCb experiment has recorded samples of charm baryons that are
larger than any previous sample by several orders of magnitude, through both prompt production and as secondary products
of $b$-hadron decays. Given the large deviation seen in the recent $\Omegac$ lifetime
measurement, the tension in the $\Lc$ lifetime measurements, and the overall relatively poor precision on the $\Lc$, $\Xicp$ and $\Xicz$ lifetimes compared 
to those for the charm mesons, it is important to have additional precise measurements of the lifetimes of these baryons.

This paper reports new measurements of the lifetimes of the $\Lc$, $\Xicp$ and $\Xicz$ baryons using samples of
semileptonic $\Lb\to\Lc\mun\neumb X$, $\Xibz\to\Xicp\mun\neumb X$, and $\Xibm\to\Xicz\mun\neumb X$ decays, 
respectively.\footnote{Throughout the text, charge-conjugate processes are implicitly included.}
The symbol $X$ represents any additional undetected particles. 
The $\Lc$ and $\Xicp$ baryons are both reconstructed in the $p\Km\pip$ final state and the $\Xicz$ baryon is observed through its decay to $p\Km\Km\pip$.
The technique employed to measure the charm-baryon lifetimes
follows that used to measure the $\Omegac$ lifetime in Ref.~\cite{LHCb-PAPER-2018-028}.

To reduce the uncertainties associated with systematic effects, the lifetime ratio
\begin{align}
r_{H_c}\equiv\frac{\tau_{H_c}}{\tau_{\Dp}}
\label{eq:rtau}
\end{align}
\noindent is measured, where the $\Dp$ meson is reconstructed using $B\to\Dp\mun\neumb X$ decays, with \mbox{$\Dp\to\Km\pip\pip$}. 
The symbols $H_b$ and $H_c$ are used here and throughout to refer to the $b$ or $c$ hadron in any of the 
modes indicated above.

The measurements presented in this paper use proton-proton ($pp$) collision data samples collected by the LHCb experiment,
corresponding to an integrated luminosity of 3.0\invfb, of which 1.0\invfb was recorded at 
a center-of-mass energy of 7\tev and 2.0\invfb at 8\tev.
The \lhcb detector~\cite{Alves:2008zz,LHCb-DP-2014-002} is a single-arm forward
spectrometer covering the \mbox{pseudorapidity} range $2<\eta <5$,
designed for the study of particles containing \bquark or \cquark
quarks. The tracking system provides a measurement of the momentum, \ptot, of charged particles with
a relative uncertainty that varies from 0.5\% at low momentum to 1.0\% at 200\gevc.
The minimum distance of a track to a primary vertex (PV), the impact parameter (IP), 
is measured with a resolution of $\sigma_{\rm IP}=(15+29/\pt)\mum$~\cite{LHCb-DP-2014-001},
where \pt is the component of the momentum transverse to the beam, in\,\gevc.
Charged hadrons are identified using information
from two ring-imaging Cherenkov (RICH) detectors~\cite{LHCb-DP-2012-003}.
Muons are identified by a
system composed of alternating layers of iron and multiwire
proportional chambers~\cite{LHCb-DP-2013-001}. The online event selection is performed by a trigger~\cite{LHCb-DP-2012-004}, 
which consists of a hardware stage, based on information from the calorimeter and muon
systems, followed by a software stage, which applies a full event reconstruction.

Simulation is required to model the effects of the detector acceptance and resolution,
as well as the imposed selection requirements. Proton-proton collisions are simulated using
\pythia~\cite{Sjostrand:2006za,*Sjostrand:2007gs} with a specific \lhcb
configuration~\cite{LHCb-PROC-2010-056}.  Decays of hadronic particles
are described by \evtgen~\cite{Lange:2001uf}, in which final-state
radiation is generated using \photos~\cite{Golonka:2005pn}. The
interaction of the generated particles with the detector and its
response are implemented using the \geant toolkit~\cite{Allison:2006ve, *Agostinelli:2002hh} as described in
Ref.~\cite{LHCb-PROC-2011-006}. 

Samples of candidate semileptonic $H_b$ decays are formed by combining a $\mun$ candidate with a
charm-hadron candidate, reconstructed through one of the following modes:
$\Lc\to p\Km\pip$, $\Xicp\to p\Km\pip$, $\Xicz\to p\Km\Km\pip$, or $\Dp\to\Km\pip\pip$.
All final-state charged particles are required to be detached from all PVs in the event. This selection is
based upon a quantity $\chisqip$, which is the difference in the $\chisq$ of the PV fit with and 
without the inclusion of the particle under consideration. The requirement on $\chisqip$ 
for the $p$, $\Km$ and $\pip$ ($\mun$) candidates corresponds to about 2$\sigma_{\rm IP}$ (3$\sigma_{\rm IP}$).
The muon is required to have $\pt>1\gevc$, $p>6\gevc$ and have particle identification (PID) information  
consistent with that of a muon. The final-state hadrons must have PID information consistent with their assumed particle hypotheses, and have
$\pt>0.25\gevc$ and $p>2\gevc$.  To remove the contribution from promptly produced charm
baryons, the reconstructed trajectory of the $H_c$ candidate must not point back to any PV in the event.
Only $H_c$ candidates that have an invariant mass within 60\mevcc of their known mass are retained. 

The $H_c\mun$ combinations are required to form a good quality vertex and satisfy the invariant mass requirement, $m(H_c\mun)<8.0$\gevcc. 
Random combinations of $H_c$ and $\mun$ are suppressed by requiring the $H_c$ decay vertex to be downstream of the reconstructed $H_c\mun$ decay vertex.
In events with more than one PV, the $b$-hadron candidate and its decay products are associated to the PV for which 
the $\chisqip$ of the $b$ hadron is smallest.

The dominant source of background in the $H_b\to H_c\mun$ samples is from other semileptonic $b$-hadron decays.
To suppress the background in the $\Lc$ and $\Xicp$ samples from misidentified $\Ds\to\Kp\Km\pip$, $\Dp\to\Km\pip\pip$, $\Dstarp\to\Dz(\to\Km\pip)\pip$, and 
$D\to\phi(\to\Kp\Km)X$ decays, a set of vetoes is employed. 
The vetoes are only applied to candidates that have an invariant mass consistent (within $\sim2.5$ times the mass resolution) with 
either the known $\Dsp$ mass, $\Dp$ mass, the $\Dstarp-\Dz$ mass difference, or the $\phi$ meson mass, after substituting either the kaon or pion mass 
in place of the proton mass in the reconstructed decay chain. For those candidates, tighter PID requirements are imposed such that any
peaking contribution is removed. The veto removes about 1--2\% of the signal, and reduces the total background by
about 15\% (25\%) in the $\Lc$ ($\Xicp$) samples.
Potential contamination in the $\Xicz$ sample from fully reconstructed, but misidentified, four-body $\Dz$ meson decays 
has been investigated, and is found to be negligible. After all selections, the dominant source of background is from
real muons combined with partially reconstructed or misidentified charm-hadron decays.

After applying the above selections, the $\Xicp$ sample still has a lower signal-to-background ratio than the $\Lc$ and 
$\Xicz$ samples.
To improve the signal-to-background ratio in the $\Xicp\mun$ sample, a boosted decision tree (BDT) 
discriminant~\cite{Breiman, AdaBoost} is built from 18 variables. The variables are the $\chisq$ for the $\Xibz$ and $\Xicp$ 
decay-vertex fits, and for each final-state hadron: $p$, $\pt$, $\chisqip$ to the associated PV, and a PID response variable.
The BDT is trained using simulated 
$\Xibz\to\Xicp\mun\neumb X$ decays for the signal, while background is taken from the $\Xicp$ mass sidebands, 
${30<|m(p\Km\pip)-m_{\Xicp}|<50\mevcc}$,
where $m_{\Xicp}$ is the known $\Xicp$ mass~\cite{PDG2018}. Only a loose requirement on the BDT is employed, which provides an
efficiency of about 97\% for signal decays while suppressing 40\% of the background.

Signal candidates must satisfy a well-defined set of 
hardware and software trigger requirements. At the hardware level, signal candidates are required to
include a high $\pt$ muon. At the software level, they must pass a topological multivariate selection 
designed to provide an enriched sample of beauty hadrons decaying to multibody final states containing a muon~\cite{BBDT}.

\begin{figure}[tb]
\centering
\includegraphics[width=0.48\textwidth]{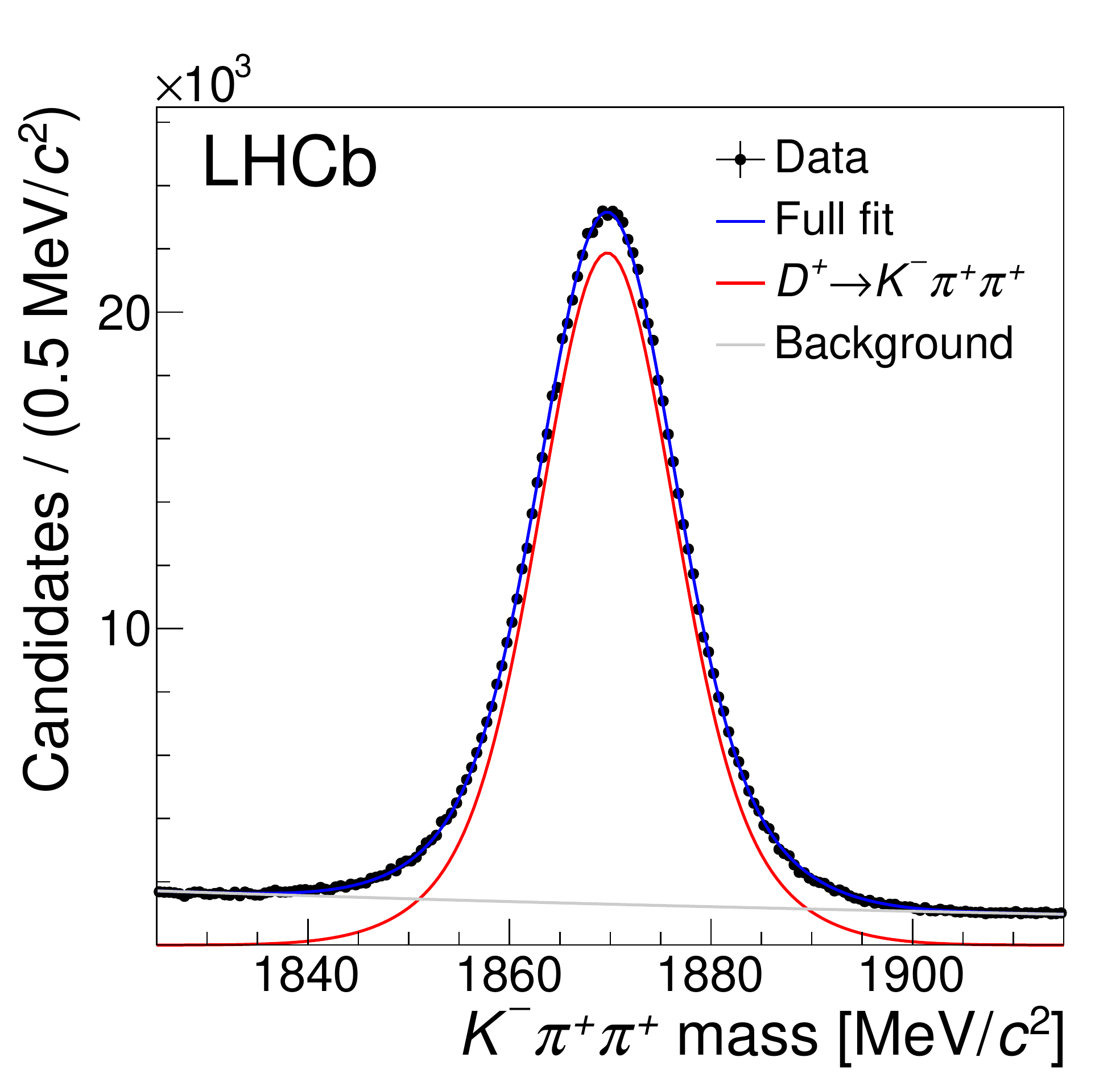}
\includegraphics[width=0.48\textwidth]{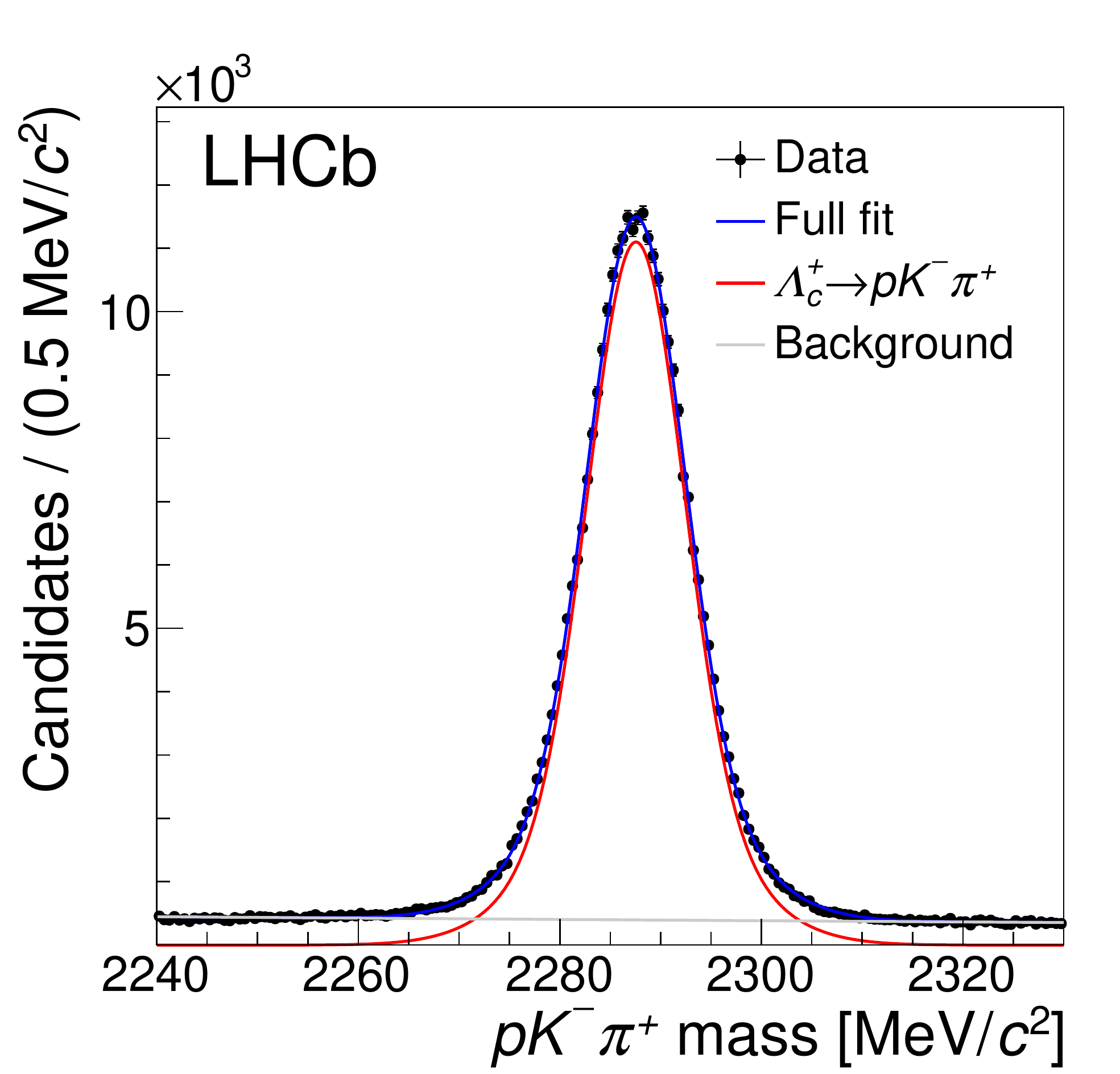}
\includegraphics[width=0.48\textwidth]{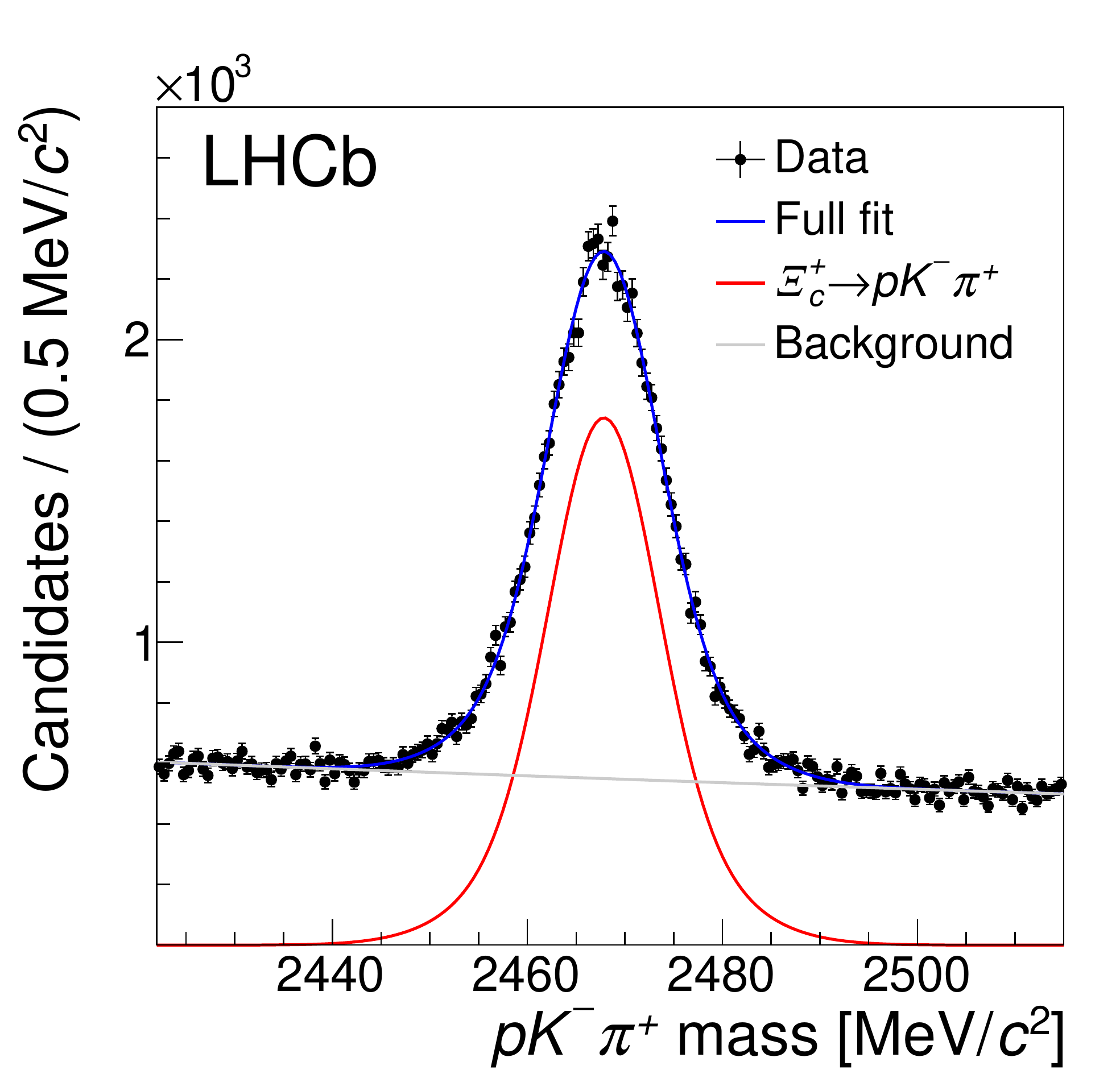}
\includegraphics[width=0.48\textwidth]{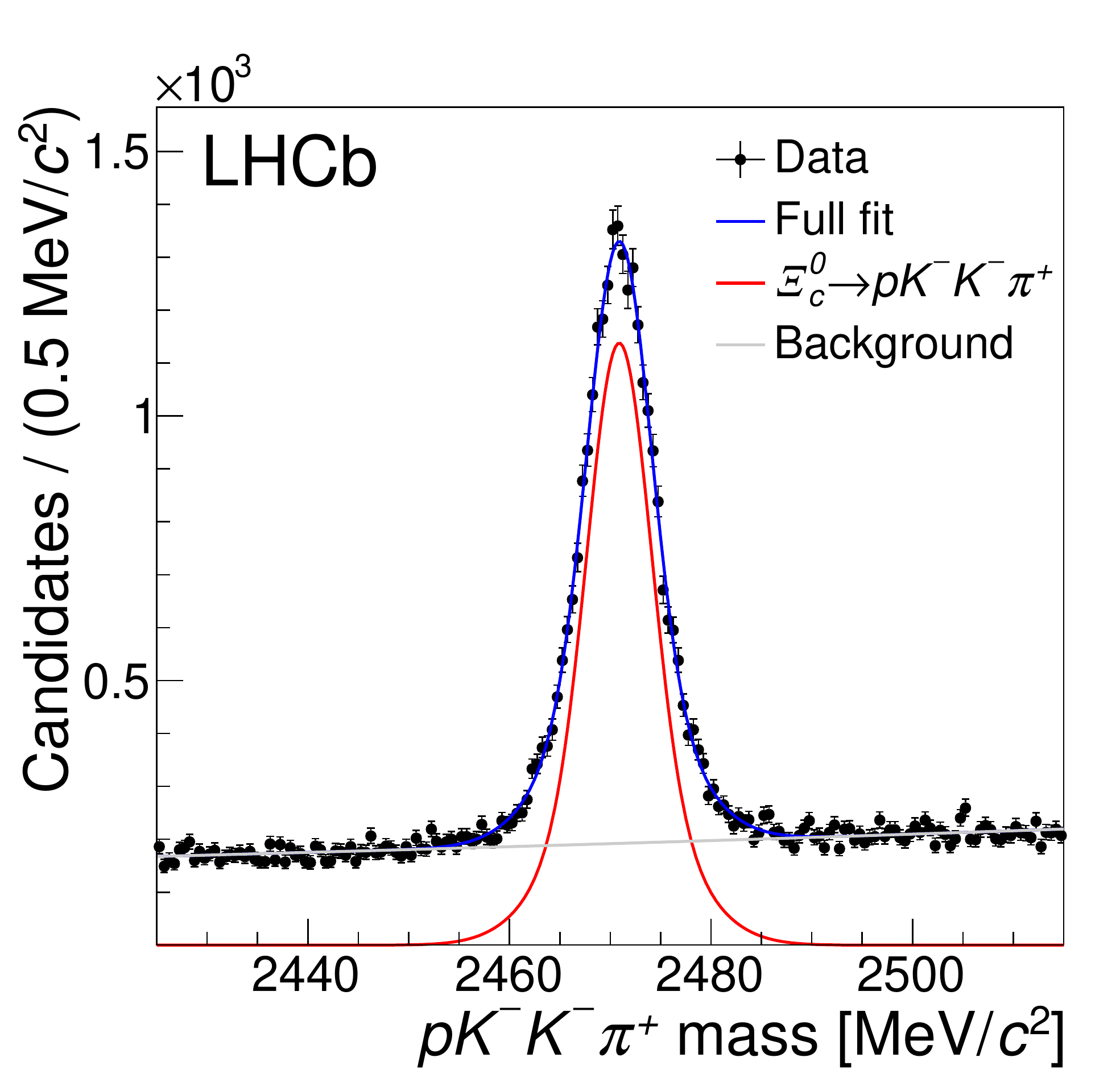}
\caption{\small{Invariant-mass distributions for candidate (top left) $\Dp$ in $B\to\Dp\mun\neumb X$,
(top right) $\Lc$ in $\Lb\to\Lc\mun\neumb X$,
(bottom left) $\Xicp$ in ${\Xibz\to\Xicp\mun\neumb X}$, and
(bottom right) $\Xicz$ in ${\Xibm\to\Xicz\mun\neumb X}$ candidate decays.
The results of the fits, as described in the text, are overlaid.}}
\label{fig:MassPlots}
\end{figure}

The invariant-mass distributions for the selected $\Dp$, $\Lc$, $\Xicp$ and $\Xicz$ candidates in the $H_c\mun$ final states 
are shown in Fig.~\ref{fig:MassPlots}. For the $\Lc$ and $\Dp$ samples only a 10\% randomly selected sub-sample of events is used 
in this analysis, since the full yield is much larger than is needed in this analysis given the anticipated size of the systematic 
uncertainties. A binned maximum-likelihood fit is performed to each of the four samples to obtain the signal yields.
For each mass distribution, the signal shape is parametrized as the sum of two Gaussian functions with a common mean, and the 
background shape is described using an exponential function. All signal and background shape parameters are freely varied in the fit. 
The resulting signal yields are given in Table~\ref{tab:yields}. The $\Xicp$ and $\Xicz$ yields are
about 100 times larger than any previous sample used to measure the lifetimes of these baryons, and the $\Lc$ sample is about 40 times
larger.

\begin{table*}[b]
\begin{center}
\caption{\small{Yields from the binned maximum-likelihood fits to the $H_c$ invariant mass spectra in $H_c\mun$ signal candidates. For the
$\Lc$ and $\Dp$ modes, only 10\% of the sample is used, since the yields in the full data set are much larger than needed in this analysis.}}
\begin{tabular}{lc}
\hline\hline
\\ [-2.5ex]
$H_c$                         &  Yield ($10^3$) \\
\hline  
$\Dp$                           &  $809.4\pm1.3$ \\
$\Lc$                           &  $303.5\pm0.7$ \\
$\Xicp$                         &  $~55.8\pm0.5$ \\
$\Xicz$                         &  $~21.6\pm0.2$ \\
\hline\hline
\end{tabular}
\label{tab:yields}
\end{center}
\end{table*}

The decay time of each $H_c$ candidate is determined from the positions of the $H_b$ and $H_c$ decay vertices, and the measured 
$H_c$ momentum. Because $b$ hadrons have a mean lifetime of about 1.5~ps, the decay vertices are well separated from the PV.
As a result, systematic effects due to lifetime-biasing selections in the trigger or offline analysis are greatly reduced compared to
promptly produced charmed baryons. 

The background-subtracted decay-time spectra are obtained using the \sPlot technique~\cite{Pivk:2004ty}, where the 
measured $H_c$ mass is used as the discriminating variable. To improve the accuracy of the \sPlot background subtraction, a correction
to the $H_c$ mass is applied to remove a small dependence of the mean reconstructed $H_c$ mass on its reconstructed decay time, $t_{\rm rec}$.
This correction is obtained by first fitting for the peak position of the reconstructed mass, $M_{\rm peak}^{H_c}(t_{\rm rec})$, in bins of
reconstructed decay time, followed by a fit for the dependence of $M_{\rm peak}^{H_c}(t_{\rm rec})$ on $t_{\rm rec}$, using the functional form
\begin{align}
M^{H_c}(0)+A[1-\exp(-t_{\rm rec}/C)].
\end{align}
The second term represents the deviation from a constant value, and is used to correct the measured $H_c$ mass of every candidate used in the \sPlot. 
For the four modes under study, the values of $A$ and $C$ range from 2.7--4.1\mevcc and 0.06--0.17~ps, respectively. 
The uncertainties in the signal yields reflect both the finite signal yield and the statistical uncertainty associated with the background subtraction.

Potential backgrounds from random $H_c\mun$ combinations, where the muon is not produced directly at the $H_b$ decay vertex, 
could lead to a bias on the lifetime. Such decays include $H_b\to H_c\taum\neutb$, \mbox{$\taum\to\mun\neut\neumb$} and 
$H_b\to H_c\Dbar,~\Dbar\to\mun X$, where $\Dbar$ represents a $\Dsm$, $\Dm$ or $\Dzb$ meson.
These backgrounds are a small fraction of the observed signal, 
about 3\% in total, and have decay-time spectra that are similar to the genuine $H_c\mun\neumb$ final state due to the 
$\chisq$ requirements on the $H_b$ vertex fit. The effect of these backgrounds is studied with simulation
and pseudoexperiments, and is included as a source of systematic uncertainty. 

\begin{figure}[tb]
\centering
\includegraphics[width=0.48\textwidth]{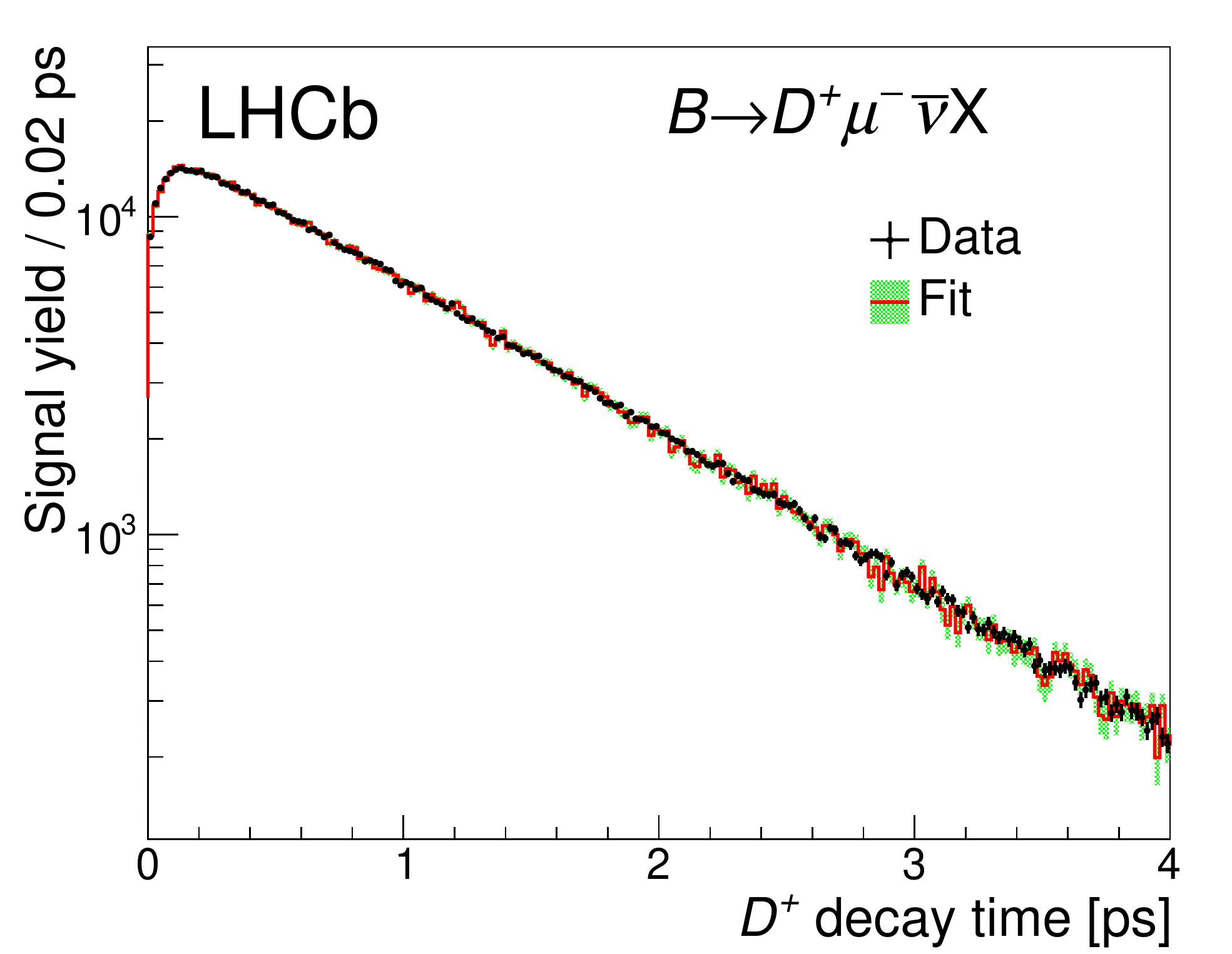}
\includegraphics[width=0.48\textwidth]{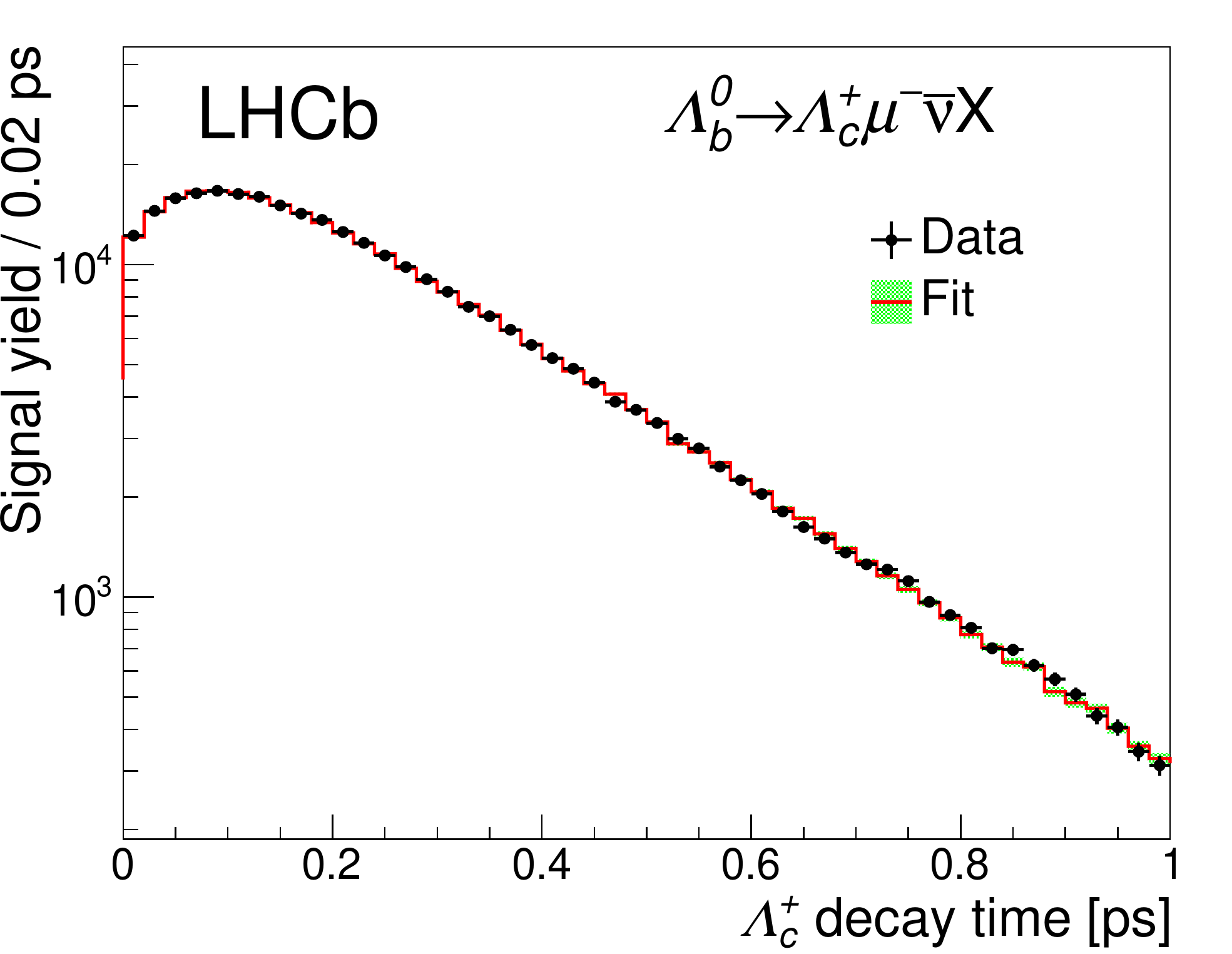}
\includegraphics[width=0.48\textwidth]{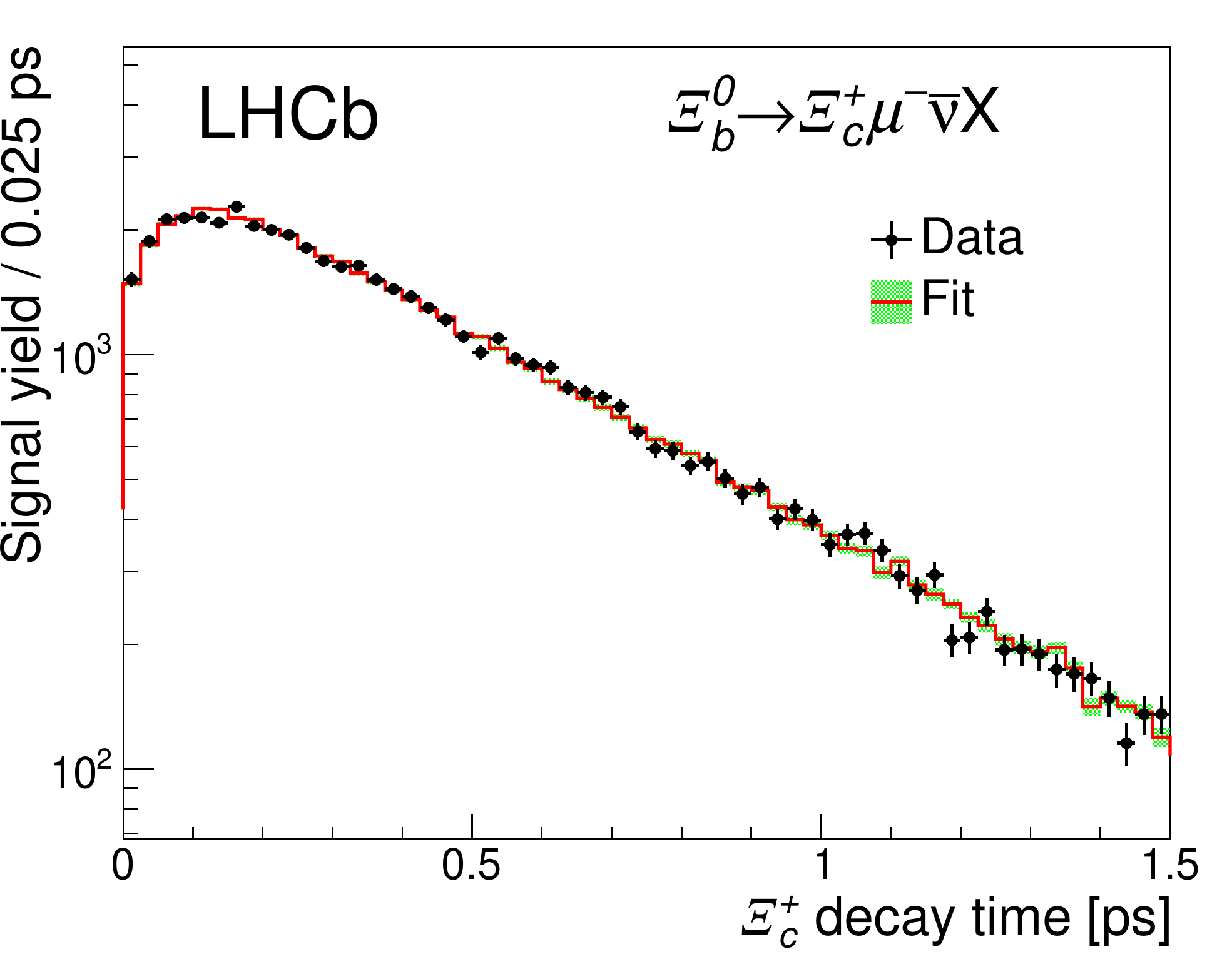}
\includegraphics[width=0.48\textwidth]{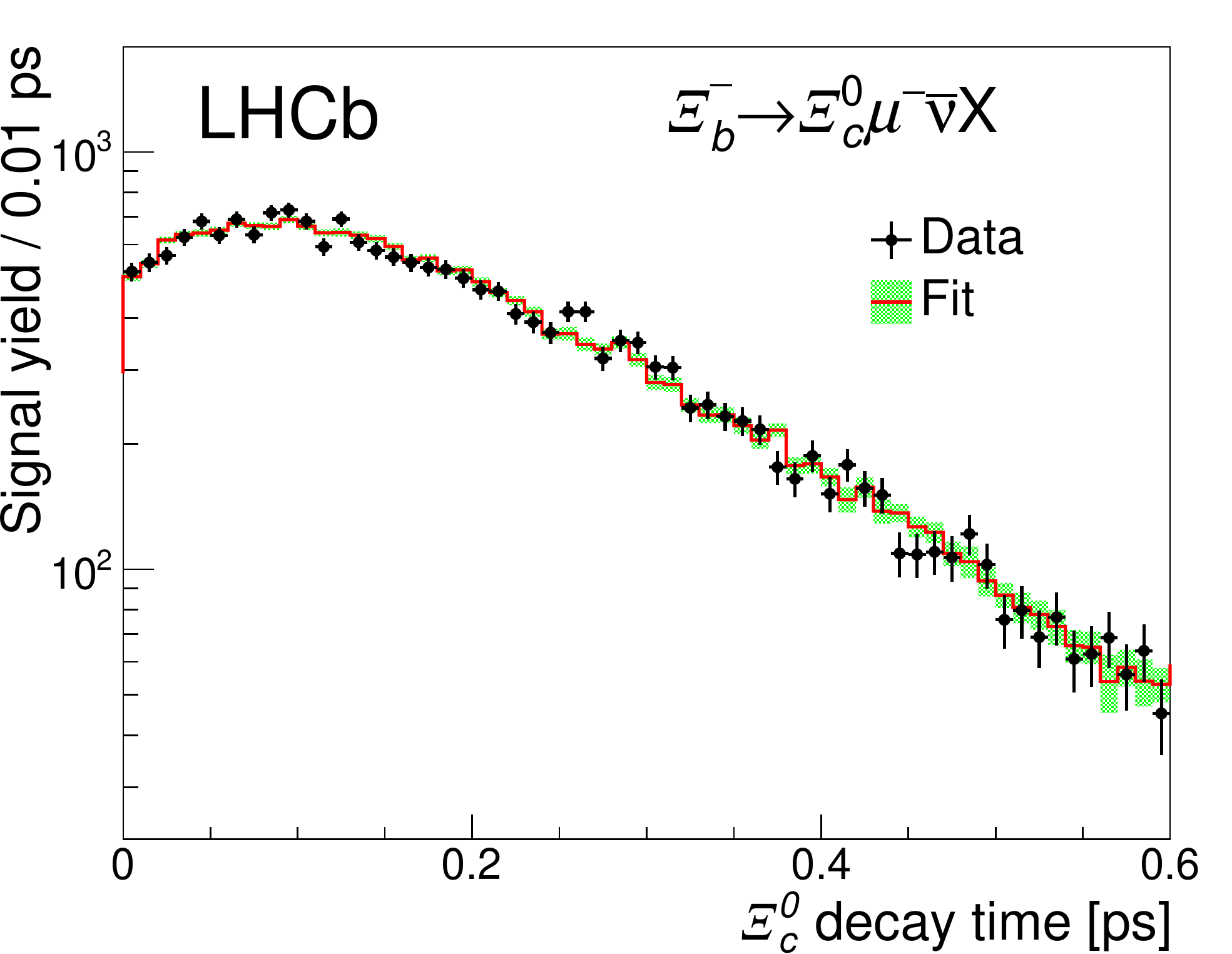}
\caption{\small{Decay-time spectra for (top left) $\Dp$ signal in $B\to\Dp\mun\neumb X$, (top right) $\Lc$ signal in
$\Lb\to\Lc\mun\neumb X$, (bottom left) $\Xicp$ signal in $\Xibz\to\Xicp\mun\neumb X$, and 
(bottom right) $\Xicz$ signal in $\Xibm\to\Xicz\mun\neumb X$ candidate decays.
Overlaid are the fit results, as described in the text, along with the uncertainties due to finite sizes of the simulated samples.}}
\label{fig:XcDecaytimeRatioFits}
\end{figure}

The decay-time spectra for the $\Dp$, $\Lc$, $\Xicp$ and $\Xicz$ signals are shown in Fig.~\ref{fig:XcDecaytimeRatioFits}, along 
with the results of the fits described below. Only $H_c$ candidates with decay time larger than zero are used in the fit.
The decrease in signal yield as the decay time approaches zero is mainly due to 
the $H_c$ decay-time resolution, typically in the 85--100~fs range, which results in migration of the signal into the negative 
decay-time region.

The charm-hadron lifetimes are determined by fitting the decay-time spectra using a binned $\chi^2$ fit over the
ranges shown in Fig.~\ref{fig:XcDecaytimeRatioFits}. The signal decay-time model takes the form
\begin{align}
S(t_{\rm rec};\tau_{\rm sim}^{H_c}) = f(t_{\rm rec};\tau_{\rm sim}^{H_c}) g(t_{\rm rec}) \beta(t_{\rm rec}),
\end{align}
where $f(t_{\rm rec};\tau_{\rm sim}^{H_c})$ is a signal template of reconstructed decay times obtained from the full LHCb simulation
with input lifetime $\tau_{\rm sim}^{H_c}$. The selection requirements applied to the simulation are identical to those applied to the
data. The function
\begin{align}
g(t_{\rm rec})=\exp(-t_{\rm rec}/\tau_{\rm fit}^{H_c}) / \exp(-t_{\rm rec}/\tau_{\rm sim}^{H_c})
\label{eq:gt}
\end{align}
\noindent weights the simulated template with lifetime $\tau_{\rm sim}^{H_c}$ to a lifetime value
$\tau_{\rm fit}^{H_c}$. Because the weighting function $g(t_{\rm rec})$ depends on 
the reconstructed decay time, $t_{\rm rec}$, rather than the true decay time, there is a dependence of 
$S(t_{\rm rec};\tau_{\rm sim}^{H_c})$ on the $\tau_{\rm sim}^{H_c}$ value used to generate the template.
The simulation uses the known $\Dp$ lifetime, $\tau_{\rm sim}^{\Dp}=1040$~fs, which is accurately measured
by many experiments~\cite{PDG2018}. 
Since the charm-baryon lifetimes in this analysis are expected to have a better precision 
than the existing world average values, a number of different $\tau_{\rm sim}^{H_c}$ templates are produced.
An optimization procedure, as described below, is used to determine the
best choice of $\tau_{\rm sim}^{H_c}$ to use for the charm-baryon templates.
The simulation includes contributions from $H_c\taum\neutb X$ final states as well as excited charm hadrons.

The function $\beta(t_{\rm rec})$ corrects for a small difference in the efficiency
between data and simulation for reconstructing tracks in the vertex detector that originate far from the beamline~\cite{LHCb-PAPER-2013-065}. 
As discussed in Ref.~\cite{LHCb-PAPER-2018-028}, $\beta(t)$ is calibrated using the precisely known value of the $\Dp$ lifetime,
from which it is found that $\beta(t_{\rm rec}) = 1+\beta_0 t_{\rm rec}$, with
$\beta_0 = (-0.89\pm0.32)\times10^{-2}$~ps$^{-1}$.  The result of the binned $\chi^2$ fit to the $\Dp$ decay-time spectrum 
after this correction is applied is shown in Fig.~\ref{fig:XcDecaytimeRatioFits}\,(top left), where the fitted lifetime is 
found to be ${\tau_{\rm fit}^{\Dp}=1042.0\pm1.7\stat~\rm{fs}}$. The inclusion of the $\beta(t)$ term in $S(t_{\rm rec})$ amounts to about a 
1\% positive correction to the measured lifetime.

Since beauty baryonic decays are not perfectly described by the simulation, the simulated events are weighted in bins of
$(\pt,\eta)$ of the beauty baryon and the mass $m(H_c\mun)$ of the $H_c\mun$ system to match that which is
observed in background-subtracted data. The simulation is also weighted to match all of the
two-body invariant mass projections among the $H_c$ decay products. After all of these weights are applied, excellent agreement
is seen for a wide range of observables in these decays, most notably those that are used in the BDT. These 
weights are applied in the formation of the $f(t_{\rm rec};\tau_{\rm sim}^{H_c})$ templates. 

The lifetime of each charmed baryon is determined from a simultaneous fit to its decay-time spectrum and that
of the $\Dp$ meson. In these fits, $\tau_{\rm fit}^{H_c}$ in Eq.~\eqref{eq:gt} is replaced by $r_{H_c}\tau_{\rm fit}^{\Dp}$
in order to reduce systematic uncertainties. Thus,
the free parameters in the fit are $r_{H_c}$, as shown in Eq.~\eqref{eq:rtau}, and $\tau_{\rm fit}^{\Dp}$. 
In the $\Xicz$ decay-time fit, $\beta_0$ is scaled by 4/3 since the effect scales with the number of charged 
final-state particles in the $H_c$ decay~\cite{LHCb-PAPER-2013-065}.

The procedure for determining the optimal values of $\tau_{\rm sim}^{H_c}$ to use in forming the templates $f(t_{\rm rec};\tau_{\rm sim}^{H_c})$ 
is first developed and validated using simulation. A series of templates, $f_i(t_{\rm rec};\tau_{\rm sim}^{H_c})$, spanning a wide range of 
$\tau_{\rm sim}^{H_c}$ values is produced for
each charm baryon. From one template with true lifetime $\tau_{\rm sim}^{H_c,\rm true}$, a pseudo-dataset set of decay times is formed that has comparable yield to 
that of the data. The decay-time fit is then performed using each template $f_i(t_{\rm rec};\tau_{\rm sim}^{H_c})$ to this pseudo-dataset, with each fit yielding a value of 
$\tau_{\rm fit}$ and a $\chi^2$ of the fit. Examination of the results show, as expected, that when the pseudo-dataset are fit using the correct template, 
$\tau_{\rm fit}$ is consistent with $\tau_{\rm sim}^{H_c, \rm true}$. 
Conversely, when the same pseudo-dataset is fit with an alternate template that is produced with a significantly different input 
value of $\tau_{\rm sim}^{H_c}$, $\tau_{\rm fit}$ deviates from $\tau_{\rm sim}^{H_c, \rm true}$. Thus the criterion for choosing the optimal template is to 
select that in which $\tau_{\rm fit}$ is closest to $\tau_{\rm sim}^{H_c}$. The chosen template is also found to have the lowest fit $\chi^2$, 
which provides additional support for the method of determining the optimal template. 

Applying this same criterion to the data, the optimal values of $\tau_{\rm sim}^{H_c}$ are found to be $\tau_{\rm sim}^{\Lc}=203$~fs, $\tau_{\rm sim}^{\Xicp}=455$~fs, 
and $\tau_{\rm sim}^{\Xicz}=155$~fs. As with the pseudo-data, these optimal values also yield the lowest $\chi^2$ value for the
decay-time fit. Slightly different values of $\tau_{\rm sim}^{H_c}$ are not excluded by the procedure, and are considered as a source of
systematic uncertainty.

The results of the fits to the $\Lc$, $\Xicp$ and $\Xicz$ decay-time distributions using the best-fit templates are 
shown in Fig.~\ref{fig:XcDecaytimeRatioFits}\, and corresponds to the ratios
\begin{align*}
  r_{\Lc} &= 0.1956\pm0.0010, \\
  r_{\Xicp} &= 0.4392\pm0.0034, \\
  r_{\Xicz} &= 0.1485\pm0.0017,
\end{align*}
\noindent where the uncertainties are statistical only.
Multiplying these ratios by the $\Dp$ lifetime~\cite{PDG2018}, leads to the lifetimes
\begin{align*}
  \tau_{\Lc} &= 203.5\pm1.0~{\rm fs}, \\
  \tau_{\Xicp} &= 456.8\pm3.5~{\rm fs}, \\
  \tau_{\Xicz} &= 154.5\pm1.7~{\rm fs}.
\end{align*}
\noindent The statistical precision of these measurements is 
5--8 times better than those of the current world average values~\cite{PDG2018}. 

A number of sources of systematic uncertainty on the measured ratios $r_{H_c}$ are summarized in Table~\ref{tab:CharmSyst}.
The decay-time acceptance correction, $\beta(t_{\rm rec})$, leads to an uncertainty of 0.5\% on $r_{H_c}$. This uncertainty includes a contribution from 
the finite $B\to\Dp\mun X$ sample sizes and the choice of fit function. 

The technique for finding the correct template is based on choosing that in which the fitted lifetime is most consistent with the
value used in the simulation. The uncertainty due to this choice is estimated by repeating the decay-time fit using
alternative templates that have simulated lifetimes that differ from the nominal one by 
two times the uncertainty on the fitted lifetime. The difference between the fitted values of $r_{H_c}$ for these 
alternative templates and the nominal one is assigned as a systematic uncertainty.

The $\Lb$, $\Xibz$ and $\Xibm$ lifetimes are not known precisely, and this has a small effect on the decay-time
acceptance. To study this effect, simulated decays are weighted to produce either a shorter or longer $H_b$ lifetime,
based on the known uncertainties on the $b$-baryon lifetimes~\cite{PDG2018}. New signal templates are formed, 
and the fits are repeated. The change in the fitted value of $r_{H_c}$ is assigned as a systematic uncertainty.

Studies of the $\Dp$ calibration mode show a small difference in the reconstruction efficiency between data and simulation,
which is described by the $\beta_0$ parameter. This parameter has a small dependence on the $\pt$ and $\eta$ of the $H_b$ hadron. 
While the signal mode simulations are weighted to match the ($\pt$, $\eta$) spectrum observed in data, the weighting is imperfect.
A difference would lead to a small bias in the average value of $\beta_0$. 
The uncertainty on $r_{H_c}$ is obtained by taking into account the variation of $\beta_0$ in different 
$\pt$ and $\eta$ ranges, and the extent to which the ($\pt$, $\eta$) spectrum differs between data and simulation
for each of the decay modes.

The decay-time resolution is checked by comparing the $\Dz$ decay-time spectra in
$\Bm\to\Dz\pim$ decays between data and simulation, where no explicit requirement on the $\Dz$ flight distance is applied.
The simulation reproduces the data well. A second check is performed where the $\Lc$ lifetime 
is fitted using a template that is produced with an additional smearing which increases the decay-time resolution by 2.5\%. 
The change increases the fit $\chisq$ substantially,
with only a small change of 0.3~fs in the fitted lifetime. This difference is considered negligible, and no systematic
uncertainty due to modeling the decay-time resolution is assigned.

The method for background subtraction uses the \sPlot technique, which relies on a specific choice for modeling the 
signal and background distributions in the charm-hadron invariant-mass spectra. To quantify a possible systematic effect on $r_{H_c}$,
the decay-time spectra in data are obtained using a different background-subtraction technique. Instead of the \sPlot method,
signal and sideband regions are defined for each of the mass spectra, and for each charm baryon the decay-time spectrum of candidates
from the sideband regions are subtracted from the spectrum obtained from the signal region. The resulting background-subtracted
decay-time spectra are then fitted using the decay-time fit described previously. The difference between this result and the nominal 
one is assigned as a systematic uncertainty. 

The decay-time spectra in the $H_c\mun$ samples have small contributions from random combinations of
$H_c$ and $\mun$ candidates [($0.8\pm0.2)$\% of the signal], as well as backgrounds where the muon 
comes from either a $\taum$ [($1.8\pm0.3$)\%] or a semileptonic $D$ decay [($0.5\pm0.2)$\%]. The impact of these 
backgrounds is assessed using pseudoexperiments, as described in Ref.~\cite{LHCb-PAPER-2018-028}.

The systematic uncertainty due to the finite size of the simulated samples used to produce the signal templates
is assessed by repeating the fit to the data many times, where in each fit the
simulated-template bin contents are fluctuated within their uncertainties. The standard deviation of the distribution of the
fitted $r_{H_c}$ values is assigned as a systematic uncertainty.  The total systematic uncertainty on $r_{H_c}$ is about 0.6\% for
the $\Lc$ and $\Xicp$ measurements, and about 1.2\% for that of the $\Xicz$ baryon.

\begin{table*}[tb]
\begin{center}
\caption{\small{Summary of systematic uncertainties on the ratio of the charm baryon to $\Dp$ meson lifetimes (in units of $10^{-4}$).
The statistical uncertainty on the measurements is also provided for reference.}}
\begin{tabular}{lccc}
\hline\hline
Source                         & $r_{\Lc}$ & $r_{\Xicp}$ & $r_{\Xicz}$ \\
\\ [-2.5ex]
\hline          
Decay-time acceptance                     &   6     &      13      &      4         \\
$H_c$ lifetime                            &   4     &       4      &      12         \\
$H_b$ lifetime                            &   1      &      3      &      0         \\
$H_b$ production spectra                  &    2     &      4      &      1         \\
Background subtraction                    &    8     &      17     &      7         \\
$H_c(\taum,~D$, random $\mun$)            &    5     &      11     &      3         \\
Simulated sample size                     &    4     &      13     &       5        \\
\hline
Total systematic                          &    13     &     28     &     16      \\
\hline
Statistical uncertainty                   &    10     &     34     &     17        \\
\hline\hline
\end{tabular}
\label{tab:CharmSyst}
\end{center}
\end{table*}

In summary, $pp$ collision data samples at 7\tev and 8\tev center-of-mass energies collected by the LHCb experiment,
corresponding to 3.0\invfb of integrated luminosity, are used to measure the lifetimes of the $\Lc$, $\Xicp$ and $\Xicz$ baryons. 
For the $\Lc$ and $\Dp$ samples, only 10\% of the integrated luminosity is used for this measurement.
The lifetimes, measured relative to that of the $\Dp$ meson, are determined to be
\begin{align*}
  r_{\Lc} &= 0.1956\pm0.0010\pm0.0013, \\
  r_{\Xicp} &= 0.4392\pm0.0034\pm0.0028, \\
  r_{\Xicz} &= 0.1485\pm0.0017\pm0.0016,
\end{align*}
\noindent where the first uncertainty is statistical and the second is systematic. After multiplying by the
known $\Dp$ lifetime of $1040\pm7$~fs~\cite{PDG2018}, the charm-baryon lifetimes are measured to be
\begin{align*}
  \tau_{\Lc} &= 203.5\pm1.0\pm1.3\pm1.4~{\rm fs}, \\
  \tau_{\Xicp} &= 456.8\pm3.5\pm2.9\pm3.1~{\rm fs}, \\
  \tau_{\Xicz} &= 154.5\pm1.7\pm1.6\pm1.0~{\rm fs},
\end{align*}
\noindent where the last uncertainty is due to the uncertainty in the $\Dp$ lifetime. 
The $\Lc$ and $\Xicp$ lifetimes are measured with about 1\% precision and are consistent with the existing world averages.
The $\Xicz$ lifetime is measured with about 1.8\% precision, and is $3.3\sigma$ larger than the world average value of 
$112^{+13}_{-10}$~fs. These measurements have uncertainties that are approximately 3--4 times smaller than those of the existing 
world average values, and have precision comparable to that achieved for charm mesons.

\section*{Acknowledgements}
%
%
\noindent 
We express our gratitude to our colleagues in the CERN
accelerator departments for the excellent performance of the LHC. We
thank the technical and administrative staff at the LHCb
institutes. We acknowledge support from CERN and from the national agencies:
CAPES, CNPq, FAPERJ and FINEP (Brazil); 
MOST and NSFC (China); CNRS/IN2P3 (France); 
BMBF, DFG and MPG (Germany); 
INFN (Italy); 
NWO (Netherlands); 
MNiSW and NCN (Poland); 
MEN/IFA (Romania); 
MSHE (Russia); 
MinECo (Spain); 
SNSF and SER (Switzerland); 
NASU (Ukraine); 
STFC (United Kingdom); 
DOE NP and NSF (USA).
We acknowledge the computing resources that are provided by CERN, IN2P3
(France), KIT and DESY (Germany), INFN (Italy), SURF (Netherlands),
PIC (Spain), GridPP (United Kingdom), RRCKI and Yandex
LLC (Russia), CSCS (Switzerland), IFIN-HH (Romania), CBPF (Brazil),
PL-GRID (Poland) and OSC (USA).
We are indebted to the communities behind the multiple open-source
software packages on which we depend.
Individual groups or members have received support from
AvH Foundation (Germany);
EPLANET, Marie Sk\l{}odowska-Curie Actions and ERC (European Union);
ANR, Labex P2IO and OCEVU, and R\'{e}gion Auvergne-Rh\^{o}ne-Alpes (France);
Key Research Program of Frontier Sciences of CAS, CAS PIFI, and the Thousand Talents Program (China);
RFBR, RSF and Yandex LLC (Russia);
GVA, XuntaGal and GENCAT (Spain);
the Royal Society
and the Leverhulme Trust (United Kingdom).

\clearpage
\newpage
\addcontentsline{toc}{section}{References}
\bibliographystyle{LHCb}
\bibliography{main,standard,LHCb-PAPER,LHCb-CONF,LHCb-DP,LHCb-TDR,steve}
 
\newpage
\centerline
{\large\bf LHCb collaboration}
\begin
{flushleft}
\small
R.~Aaij$^{29}$,
C.~Abell{\'a}n~Beteta$^{46}$,
B.~Adeva$^{43}$,
M.~Adinolfi$^{50}$,
C.A.~Aidala$^{77}$,
Z.~Ajaltouni$^{7}$,
S.~Akar$^{61}$,
P.~Albicocco$^{20}$,
J.~Albrecht$^{12}$,
F.~Alessio$^{44}$
% ...

{\footnotesize \it

% ...
$ ^{7}$Universit{\'e} Clermont Auvergne, CNRS/IN2P3, LPC, Clermont-Ferrand, France\\
% ...
$ ^{12}$Fakult{\"a}t Physik, Technische Universit{\"a}t Dortmund, Dortmund, Germany\\
% ...
$ ^{20}$INFN Laboratori Nazionali di Frascati, Frascati, Italy\\
% ...
$ ^{29}$Nikhef National Institute for Subatomic Physics, Amsterdam, Netherlands\\
% ...
$ ^{43}$Instituto Galego de F{\'\i}sica de Altas Enerx{\'\i}as (IGFAE), Universidade de Santiago de Compostela, Santiago de Compostela, Spain\\
$ ^{44}$European Organization for Nuclear Research (CERN), Geneva, Switzerland\\
% ...
$ ^{46}$Physik-Institut, Universit{\"a}t Z{\"u}rich, Z{\"u}rich, Switzerland\\
% ...
$ ^{50}$H.H. Wills Physics Laboratory, University of Bristol, Bristol, United Kingdom\\
% ...
$ ^{61}$University of Cincinnati, Cincinnati, OH, United States\\
% ...
$ ^{63}$Syracuse University, Syracuse, NY, United States\\
% ...
$ ^{77}$University of Michigan, Ann Arbor, United States, associated to $^{63}$\\
% ...
}
\end{flushleft}
\end{document}